\documentclass[12pt]{article}

\usepackage[dvipdfmx]{graphicx}
\usepackage[dvipdfmx]{xcolor}
\usepackage{amsmath}
\usepackage{amssymb}
\usepackage{tikz}
\usepackage{cite}
\usepackage{fancyhdr}

\def\erase#1{{}}
\def\EqArrerase#1{{}}

\makeatletter
 
 \@addtoreset{equation}{section}
\makeatother

\def\GL{{G\kern-.12em L\kern-.04em}}
\def\OSp{{O\kern-.11em S\kern-.04em p}}
\def\IOSp{{I\kern-.06em O\kern-.11em S\kern-.04em p}}
\def\MN{{M\kern-.14em N}}
\def\NM{{N\kern-.14em M}}
\def\NL{{N\kern-.14em L}}
\def\LN{{L\kern-.11em N}}
\def\ML{{M\kern-.14em L}}
\def\LM{{L\kern-.11em M}}
\def\RN{{R\kern-.11em N}}
\def\NR{{N\kern-.14em R}}
\def\RM{{R\kern-.11em M}}
\def\MR{{M\kern-.14em R}}
\def\RL{{R\kern-.11em L}}
\def\LR{{L\kern-.11em R}}
\def\RS{{R\kern-.11em S}}
\def\SR{{S\kern-.11em R}}
\def\SN{{S\kern-.11em N}}
\def\NS{{N\kern-.11em S}}
\def\SM{{S\kern-.11em M}}
\def\MS{{M\kern-.11em S}}
\def\SL{{S\kern-.11em L}}
\def\LS{{L\kern-.11em S}}
\def\sqr#1#2{{\vcenter{\hrule height.#2pt
      \hbox{\vrule width.#2pt height#1pt \kern#1pt
          \vrule width.#2pt}
      \hrule height.#2pt}}}
\def\bra0{\langle0|}
\def\ket0{|0\rangle}
\def\soeji#1_#2#3{#1_{#2}\cdots#1_{#3}}
\def\longgLRarrow{\longleftarrow\kern-3pt\relbar\kern-3pt\relbar\kern-3pt%
\longrightarrow}
\def\longLRarrow{\longleftarrow\kern-3pt\relbar\kern-3pt\longrightarrow}
\def\longLarrow{\longleftarrow\kern-3pt\relbar\kern-3pt\relbar\kern-3pt\relbar}
\def\longRarrow{\relbar\kern-3pt\relbar\kern-3pt\relbar\kern-3pt\longrightarrow}
\def\bothDer#1#2#3{%
\overset{\kern-.7em\stackrel{#1}{#2}}{\partial_{#3}}}
 
\makeatletter
 
 \@addtoreset{equation}{section}
\makeatother

\begin{document}
\thispagestyle{fancy}

\title{Bound States in Lee's Complex Ghost Model}

\author{Ichiro Oda
\footnote{Electronic address: ioda@cs.u-ryukyu.ac.jp}
\\
{\it\small
\begin{tabular}{c}
Department of Physics, Faculty of Science, University of the 
           Ryukyus,\\
           Nishihara, Okinawa 903-0213, Japan\\      
\end{tabular}
}
}
\date{}

\maketitle

\thispagestyle{fancy}

\begin{abstract}

Quantum field theories (QFTs) including fourth derivative terms such as the Lee-Wick finite QED and 
quadratic gravity have a better ultra-violet behavior compared to standard theories with second derivative
ones, but the existence of ghost with negative norm endangers unitarity. Such a ghost in general acquires 
a pair of complex conjugate masses from radiative corrections whose features are concisely described 
by the so-called Lee model. Working with the canonical operator formalism of QFTs, we investigate 
the issue of bound states in the Lee model. We find that the bound states can be created from ghosts 
as in path integral approach although there is a nontrivial complex delta function, which is a complex 
generalization of the well-known Dirac delta function. 
Finally, the problem of amelioration of the unitarity in quadratic gravity is briefly discussed.
     
\end{abstract}

\newpage
\pagestyle{plain}
\pagenumbering{arabic}


\section{Introduction}

It is well known that in higher derivative quantum field theories (QFTs), a ghost mode with negative metric (norm) appears 
in classical action either explicitly or effectively via radiative corrections. In particular, radiative corrections make
the ghost mode with negative metric transform to a complex ghost. As for such a complex ghost, there is a surprising 
conjecture by Lee and Wick \cite{LW1, LW2} that the complex ghost cannot be created from scattering processes 
of physical particles with positive metric owing to energy conservation law of positive energy, and therefore the physical
unitarity of the S-matrix is not violated.  

However, this conjecture has been recently dismissed in Refs. \cite{KK1, KK2} where the complex delta function 
appearing at each vertex in Feynman diagrams plays an important role 
and leads to the violation of the physical unitarity of the S-matrix. In this respect, it is worthwhile to recall
the difference between the physical unitarity and the total unitarity of the S-matrix: The physical unitarity of the S-matrix 
means that the probability interpretation of a theory is possible so that the presence of physical ghost particles 
carrying negative probability breaks this physical unitarity. On the other hand, 
the total unitarity of the S-matrix, i.e., $S^\dagger S = S S^\dagger = 1$ is automatically obeyed as long as 
the Hamiltonian is Hermitian. The latter unitarity guarantees that the probability of a theory is conserved 
under time development regardless of whether there are normal particles with positive metric and
ghost ones with negative metric.  

More recently, based on path integral approach, using a rather simple scalar model which is very similar to the Lee's 
complex ghost model, it is pointed out that a pair of complex ghosts forms a bound state with positive norm 
and thus leading to the confinement of ghost particles into a normal composite particle with positive norm \cite{Asorey1, Asorey2}. 
In this article, we wish to investigate a possibility that a pair of ghost particles is transformed to
a normal bound state within the framework of the canonical operator formalism developed by Kubo and 
Kugo \cite{KK1, KK2}. We will find that the bound states can be created from ghosts as in path integral approach 
although there is a nontrivial complex delta function. 

The paper is organized as follows: In the next section, we introduce the Lee model and explain the canonical operator
formalism of the model. The correlation function of a ghost composite operator is derived in Section 3 and the equation
of pole positions is calculated in Section 4. Finally, in Section 5 we draw our conclusion.

\section{Canonical operator formalism of Lee model}

Let us start with a classical Lagrangian density of our model treated in this article: 
\begin{eqnarray}
{\cal{L}} = {\cal{L}}^{(2)} + {\cal{L}}^{\rm{int}},
\label{Lag1}  
\end{eqnarray}
where ${\cal{L}}^{(2)}$ and ${\cal{L}}^{\rm{int}}$ denote the quadratic kinetic term and the interaction one, respectively,
whose concrete expressions read\footnote{We choose the metric convention $\eta_{\mu\nu} = \rm{diag} 
( -1, +1, +1, +1)$.}
\begin{eqnarray}
{\cal{L}}^{(2)} &=& \frac{1}{2} \left[ (\partial \varphi)^2 + M^2 \varphi^2 + (\partial \varphi^\dagger)^2 
+ M^{*2} \varphi^{\dagger 2} \right],
\nonumber\\
{\cal{L}}^{\rm{int}} &=& - \frac{f}{4} (\varphi^\dagger \varphi)^2.
\label{Lag2}  
\end{eqnarray}
Here $\varphi$ is a complex scalar field and $\varphi^\dagger$ is its Hermitian conjugate, both of which
denote complex ghost fields. $M^2$ denotes the complex squared mass of the scalar field and is assumed that 
both its real part ${\rm{Re}} \, M^2$ and imaginary one ${\rm{Im}} \, M^2$ are positive definite, i.e., ${\rm{Re}} \, M^2 > 0$
and ${\rm{Im}} \, M^2 > 0$. We will call the model defined by the above Lagrangian density the Lee model since 
although the original Lee model consists of a real scalar field of positive norm and that of negative norm together
with their mixing term \cite{Lee}, we can diagonalize the Hamiltonian by introducing the complex ghost $\varphi$ and 
its Hermitian conjugate $\varphi^\dagger$ \cite{Nakanishi1}. The imaginary part of $M^2$ may be an intrinstic
one or may be due to radiative corrections as seen in a general Lagrangian density with fourth-derivative terms.

The free field equations for $\varphi$ and $\varphi^\dagger$ from ${\cal{L}}^{(2)}$ are 
\begin{eqnarray}
(\Box - M^2) \varphi = (\Box - M^{*2}) \varphi^\dagger = 0,
\label{Free-eq}  
\end{eqnarray}
so the complex scalar fields $\varphi$ and $\varphi^\dagger$ can be canonically quantized and are
expanded as \cite{Nakanishi1}
\begin{eqnarray}
\varphi (x) &=& \int \frac{d^3 q}{\sqrt{(2 \pi)^3 2 \omega_q}} \left[ \alpha (\Vec{q}) e^{i \Vec{q} \cdot \Vec{x} 
- i \omega_q x^0} + \beta^\dagger (\Vec{q}) e^{-i \Vec{q} \cdot \Vec{x} + i \omega_q x^0} \right],
\nonumber\\
\varphi^\dagger (x) &=& \int \frac{d^3 q}{\sqrt{(2 \pi)^3 2 \omega_q^*}} \left[ \alpha^\dagger (\Vec{q}) 
e^{-i \Vec{q} \cdot \Vec{x} + i \omega_q^* x^0} + \beta (\Vec{q}) e^{i \Vec{q} \cdot \Vec{x} 
- i \omega_q^* x^0} \right],
\label{FT-ghost}  
\end{eqnarray}
where 
\begin{eqnarray}
q^\mu \equiv ( \omega_q, \Vec{q} ), \qquad
\omega_q \equiv \sqrt{\Vec{q} \,^2 + M^2}, \qquad
\omega_q^* \equiv \sqrt{\Vec{q} \,^2 + M^{*2}}.
\label{p-omega}  
\end{eqnarray}
As a feature of the system with the complex mass, creation operators $\alpha^\dagger (\Vec{q}),
\beta^\dagger (\Vec{q})$ and annihilation ones $\alpha (\Vec{q}), \beta (\Vec{q})$ obey the off-diagonal
commutation relations \cite{Nakanishi1}:
\begin{eqnarray}
&{}& [ \alpha (\Vec{p}), \beta^\dagger (\Vec{q}) ] = [ \beta (\Vec{p}), \alpha^\dagger (\Vec{q}) ] 
= - \delta^3 (p-q),   \nonumber\\
&{}& [ \alpha (\Vec{p}), \alpha^\dagger (\Vec{q}) ] = [ \beta (\Vec{p}), \beta^\dagger (\Vec{q}) ] = 0. 
\label{CRs}  
\end{eqnarray}

Next, we turn our attention to the propagators of $\varphi$ and $\varphi^\dagger$. By means of 
Eqs. (\ref{FT-ghost}), (\ref{CRs}) and the definition of the T-product, it is easy to obtain the propagator of $\varphi$
as follows \cite{KK1, KK2}:
\begin{eqnarray}
&{}& D_\varphi (x-y) \equiv \langle 0 | T \varphi(x) \varphi(y) | 0 \rangle
\nonumber\\
&{}& \equiv \langle 0 | [ \theta (x^0 - y^0) \varphi(x) \varphi(y) + \theta (y^0 - x^0) \varphi(y) \varphi(x) ] | 0 \rangle
\nonumber\\
&{}& = \int \frac{d^3 q d^3 q^\prime}{(2 \pi)^3 \sqrt{2 \omega_q \cdot 2 \omega_{q^\prime}}} 
\Big[ \theta (x^0 - y^0)  e^{i ( \Vec{q} \cdot \Vec{x} - \omega_q x^0 - \Vec{q} \,^\prime \cdot \Vec{y} 
+ \omega_{q^\prime} y^0 )}  \langle 0 | \alpha (\Vec{q}) \beta^\dagger (\Vec{q} \,^\prime) | 0 \rangle
\nonumber\\
&{}& + \theta (y^0 - x^0)  e^{i ( \Vec{q} \,^\prime \cdot \Vec{y} - \omega_{q^\prime} y^0 - \Vec{q} \cdot \Vec{x} 
+ \omega_q x^0 )}  \langle 0 | \alpha (\Vec{q} \,^\prime) \beta^\dagger (\Vec{q}) | 0 \rangle \Big] 
\nonumber\\
&{}& = - \int \frac{d^3 q}{(2 \pi)^3 2 \omega_q } 
\Big[ \theta (x^0 - y^0)  e^{i \Vec{q} \cdot (\Vec{x} - \Vec{y}) - i \omega_q ( x^0 - y^0 )}  
+ \theta (y^0 - x^0)  e^{-i \Vec{q} \cdot (\Vec{x} - \Vec{y}) + i \omega_q ( x^0 - y^0 )} \Big] 
\nonumber\\
&{}& = - \int \frac{d^3 q}{(2 \pi)^3} e^{i \Vec{q} \cdot (\Vec{x} - \Vec{y})}
\int_C \frac{d q^0}{2 \pi i} \frac{e^{- i q^0 ( x^0 - y^0 )}}{q^2 + M^2} 
\nonumber\\
&{}& \equiv \int_C \frac{d^4 q}{i (2 \pi)^4} e^{i q \cdot (x - y)} D_\varphi (q), 
\label{Prop-varphi}  
\end{eqnarray}
where we have defined
\begin{eqnarray}
D_\varphi (q) \equiv - \frac{1}{q^2 + M^2}. 
\label{D-varphi}  
\end{eqnarray}
Here the contour $C$ is the well-known Lee-Wick complex contour \cite{Lee} which starts at
$q^0 = - \infty$, passes below the left pole at $q^0 = - \omega_q$ and above the right pole at
$q^0 = \omega_q$, and then goes to the infinity $q^0 = + \infty$ in the complex $q^0$-plane,
as shown in Fig. 1.
Note that this contour is automatically derived in the canonical formalism under consideration 
without recourse to the definition of the S-matrix \cite{Nakanishi1}.

\begin{tikzpicture}[>=stealth, scale=1.2]

\draw[->] (-3,0) -- (3,0) node[right] {$\mathrm{Re}\, q^0$};
\draw[->] (0,-2) -- (0,2) node[above] {$\mathrm{Im}\, q^0$};

\fill (-2,-0.5) circle (2pt) node[left] {$-\omega_{\mathbf q}$};
\fill (2,0.5)  circle (2pt) node[right] {$+\omega_{\mathbf q}$};
\draw (0,0) node[below right] {0};
\draw[thick, blue]
  (-3,0) -- (-2.3,0)
  .. controls (-2.3,-1.2) and (-1.7,-1.2) ..
  (-1.7,0)
  -- (1.7,0)
  .. controls (1.7,1.2) and (2.3,1.2) ..
  (2.3,0)
  -- (3,0);

\node[blue] at (-1,-0.3) {$C$};

\draw[->, thick, blue] (-1,0) -- (-0.5,0);

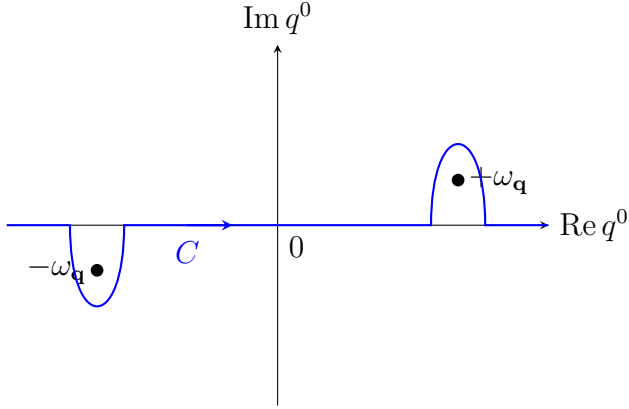
\draw  (0,-2.2) node[below] {Fig.1. The Lee-Wick contour}; 

\end{tikzpicture}

In a perfectly similar manner, the propagator of $\varphi^\dagger$ is obtained to be
\begin{eqnarray}
&{}& D_{\varphi^\dagger} (x-y) \equiv \langle 0 | T \varphi^\dagger(x) \varphi^\dagger(y) | 0 \rangle
\nonumber\\
&{}& = - \int \frac{d^3 q}{(2 \pi)^3 2 \omega_q^* } 
\Big[ \theta (x^0 - y^0)  e^{i \Vec{q} \cdot (\Vec{x} - \Vec{y}) - i \omega_q^* ( x^0 - y^0 )}  
+ \theta (y^0 - x^0)  e^{-i \Vec{q} \cdot (\Vec{x} - \Vec{y}) + i \omega_q^* ( x^0 - y^0 )} \Big] 
\nonumber\\
&{}& = - \int \frac{d^3 q}{(2 \pi)^3} e^{i \Vec{q} \cdot (\Vec{x} - \Vec{y})}
\int_R \frac{d q^0}{2 \pi i} \frac{e^{- i q^0 ( x^0 - y^0 )}}{q^2 + M^{*2}} 
\nonumber\\
&{}& \equiv \int_R \frac{d^4 q}{i (2 \pi)^4} e^{i q \cdot (x - y)} D_{\varphi^\dagger} (q), 
\label{Prop-varphi-dagger}  
\end{eqnarray}
where we have also defined
\begin{eqnarray}
D_{\varphi^\dagger} (q) \equiv - \frac{1}{q^2 + M^{*2}}. 
\label{D-varphi-dagger}  
\end{eqnarray}
Note that in this case the contour must be chosen along the real axis $R$ because of 
$M^{*2} = {\rm{Re}} \, M^2 - i {\rm{Im}} \, M^2$ with ${\rm{Im}} \, M^2 > 0$.

\section{Correlation function of composite operator}

In this section, we will derive the correlation function of a composite operator of ghost fields
on the basis of the canonical operator formalism of the Lee model as explained in the previous section. 
This problem has been recently discussed in Refs. \cite{Asorey1, Asorey2} from path integral approach. 
We will obtain the different expression of the correlation function from that of Refs. \cite{Asorey1, Asorey2}. 
It will turn out that the difference stems from a careful treatment of the Lee-Wick complex contour 
and the complex delta function.

We would like to consider a bound state problem constructed out of the following composite 
ghost operator:
\begin{eqnarray}
{\cal{O}}_{|\varphi|^2} (x) \equiv \varphi^\dagger(x) \varphi(x). 
\label{Comp-op}  
\end{eqnarray}
This is a neutral operator and is one of the most plausible operators if a bound state is made from
ghost fields. Since the other ghost composite operators are dealed with in the same way, we will focus
our attention on whether bound states are made from the operator ${\cal{O}}_{|\varphi|^2} (x)$ in Eq. (\ref{Comp-op})
or not within the framework of the canonical operator formalism.

According to the Gell-Mann-Low formula \cite{Gell-Low}, in the canonical operator formalism the correlation function 
of the ghost composite operator ${\cal{O}}_{|\varphi|^2} (x)$ is defined as\footnote{See Refs. \cite{Nishijima, 
Zimmermann1, Zimmermann2} for composite field operators.}
\begin{eqnarray}
&{}& \langle {\cal{O}}_{|\varphi|^2} (x) {\cal{O}}_{|\varphi|^2} (y) \rangle 
\equiv \langle {\bf{0}} | T [{\cal{O}}_{|\varphi|^2} (x) {\cal{O}}_{|\varphi|^2} (y) ] | {\bf{0}} \rangle 
\nonumber\\
&\equiv& \sum_{n=0}^\infty \frac{i^n}{n!} \lim_{\varepsilon \rightarrow +0} 
\int d^4 x_1 \cdots d^4 x_n \, {\rm{exp}} (- \varepsilon \sum_{i=1}^n | x_i^0 |)
\frac{1}{\langle 0 | S | 0 \rangle}
\nonumber\\
&\times& \langle 0 | T [ {\cal{L}}^{\rm{int}}(x_1) \cdots {\cal{L}}^{\rm{int}}(x_n) 
\varphi^\dagger(x) \varphi(x) \varphi^\dagger(y) \varphi(y) ] | 0 \rangle, 
\label{Corr-funct}  
\end{eqnarray}
where a generic field operator ${\bf{\Phi}}(x)$ in the interaction picture is related to $\Phi(x)$ in the
Schrodinger picture by ${\bf{\Phi}}(x) = e^{iHx^0} \Phi(0, \Vec{x}) e^{-iHx^0}$ with $H$ being the 
Hamiltonian. Moreover, $| {\bf{0}} \rangle$ is the true vacuum which is different from the vacuum of the free
theory, $| 0 \rangle$ and $S$ is the S-matrix.\footnote{$| {\bf{0}} \rangle$ is related to $| 0 \rangle$ by 
$| {\bf{0}} \rangle = e^{i \theta} U(0, -\infty)| 0 \rangle$ where $\theta$ is a phase factor and $U$ is the standard U-matrix, which
is symbolically denoted as $U(t, t_0) \equiv T \exp \left(i \int_{t_0}^t d t^\prime \, {\cal{L}}^{\rm{int}}(t^\prime)
\right)$.}
Of course, the ghost composite operator on the left hand side (LHS) of Eq. (\ref{Corr-funct})
is now defined not as the free Schrodinger operator but as the operator in the interaction picture. 

For simplicity of presentation, we will take $\varepsilon = 0$ and ${\langle 0 | S | 0 \rangle} = 1.$\footnote{$\langle 0 | S |  0 \rangle$
corresponds to the sum of vacuum polarization diagrams. The division by $\langle 0 | S |  0 \rangle$ means that one can neglect
the Feynman diagrams involving the vacuum polarization.}
Expanding the exponential up to the order $f^2$ and using the Wick theorem, we have
\begin{eqnarray}
&{}& \langle {\cal{O}}_{|\varphi|^2} (x) {\cal{O}}_{|\varphi|^2} (y) \rangle 
\nonumber\\
&=& D_{\varphi^\dagger} (x-y) D_\varphi (x-y) - i f \int d^4 z \, D_{\varphi^\dagger} (x-z) 
D_{\varphi^\dagger} (z-y) D_\varphi (x-z) D_\varphi (z-y)
\nonumber\\
&-& f^2 \int d^4 z d^4 w \, D_{\varphi^\dagger} (x-z) D_{\varphi^\dagger} (z-w) 
D_{\varphi^\dagger} (w-y) D_\varphi (x-z) D_\varphi (z-w) D_\varphi (w-y) 
\nonumber\\
&+& {\cal{O}}(f^3). 
\label{Corr-funct2}  
\end{eqnarray}
From now on, let us explicitly calculate each term on the right hand side (RHS) of this equation. As for the first term, 
$D_{\varphi^\dagger} (x-y) D_\varphi (x-y)$, paying an attention to the difference of the contours $C$ 
and $R$, the calculation proceeds as follows:
\begin{eqnarray}
&{}& D_{\varphi^\dagger} (x-y) D_\varphi (x-y)
\nonumber\\
&=& \int_R \frac{d^4 q}{i (2 \pi)^4} e^{i q \cdot (x-y)} D_{\varphi^\dagger} (q) 
\int_C \frac{d^4 q^\prime}{i (2 \pi)^4} e^{i q^\prime \cdot (x-y)} D_\varphi (q^\prime) 
\nonumber\\
&=& \int_C \frac{d^4 k}{i (2 \pi)^4} e^{i k \cdot (x-y)}  \int_R \frac{d^4 q}{i (2 \pi)^4} D_{\varphi^\dagger} (q) D_\varphi (k-q) 
\nonumber\\
&\equiv& \int_C \frac{d^4 k}{i (2 \pi)^4} e^{i k \cdot (x-y)}  G(k), 
\label{D-D}  
\end{eqnarray}
where $G(k)$ is defined by
\begin{eqnarray}
G(k) \equiv \int_R \frac{d^4 q}{i (2 \pi)^4} D_{\varphi^\dagger} (q) D_\varphi (k-q).
\label{Def-G}  
\end{eqnarray}
Note that at the second equality of Eq. (\ref{D-D}) we have made the change of variable from $q^\prime$ to $k$, which is defined as $k = q + q^\prime$,
but $q^0$ and $q^{\prime \, 0}$ run along the contours $R$ and $C$, respectively, so the variable $q^\prime$, not $q$, must be expressed 
in terms of $k$. 
  
Next, let us move on to the second term on the RHS of Eq. (\ref{Corr-funct2}). Up to an overall factor $- i f$, this term can be evaluated to be
\begin{eqnarray}
&{}& \int d^4 z \, D_{\varphi^\dagger} (x-z) D_{\varphi^\dagger} (z-y) D_\varphi (x-z) D_\varphi (z-y)
\nonumber\\
&=& \int d^4 z \, D_{\varphi^\dagger} (x-z) D_\varphi (x-z) D_{\varphi^\dagger} (z-y) D_\varphi (z-y)
\nonumber\\
&=& \int d^4 z \, \int_C \frac{d^4 k}{i (2 \pi)^4} \int_C \frac{d^4 k^\prime}{i (2 \pi)^4} e^{i ( k x - k^\prime y)} e^{-i ( k - k^\prime ) z} G(k) G(k^\prime). 
\label{O(f)}  
\end{eqnarray}
At this point, we encounter an important fact: The variables $k^0, k^{\prime \, 0}$ take values on complex numbers, so the integration
over $z^0$ gives rise to the complex delta function \cite{Nakanishi3}:  
\begin{eqnarray}
\frac{1}{(2 \pi)^4} \int d^4 z \, e^{-i ( k - k^\prime ) z} = \delta_c (k^0 - k^{\prime \, 0}) \delta^3 (k - k^\prime), 
\label{Com-delta}  
\end{eqnarray}
where the complex delta function is defined as
\begin{eqnarray}
\frac{1}{2 \pi} \int d z^0 \, e^{-i ( k^0 - k^{\prime \, 0} ) z^0} = \delta_c (k^0 - k^{\prime \, 0}), 
\label{Def-Com-delta}  
\end{eqnarray}
for general complex numbers $k^0$ and $k^{\prime \, 0}$. Thus we arrive at 
\begin{eqnarray}
&{}& \int d^4 z \, D_{\varphi^\dagger} (x-z) D_{\varphi^\dagger} (z-y) D_\varphi (x-z) D_\varphi (z-y)
\nonumber\\
&=& \int_C \frac{d^4 k d k^{\prime \, 0}}{i^2 (2 \pi)^4} e^{-i ( k^0 x^0 - k^{\prime \, 0} y^0 ) + i \Vec{k} \cdot ( \Vec{x} - \Vec{y} ) } 
\delta_c (k^0 - k^{\prime \, 0}) G(k^0, \Vec{k}) G(k^{\prime \, 0}, \Vec{k}). 
\label{O(f)-2}  
\end{eqnarray}

In a perfectly similar manner, one can calculate the third term on the RHS of Eq. (\ref{Corr-funct2}). 
Putting all terms together, we have an expression for the correlation function: 
\begin{eqnarray}
\langle {\cal{O}}_{|\varphi|^2} (x) {\cal{O}}_{|\varphi|^2} (y) \rangle 
\equiv \int_C \frac{d^4 k}{i (2 \pi)^4} \, e^{i k \cdot (x - y)} C(k),
\label{Corr-funct3}  
\end{eqnarray}
where $C(k)$ is defined by
\begin{eqnarray}
C(k) &\equiv& G(k) \Big[ 1 - f \int_C d k^{\prime \, 0} \, \delta_c (k^0 - k^{\prime \, 0}) G(k^{\prime \, 0}, \Vec{k})
\nonumber\\
&+& f^2 \int_C d k^{\prime \, 0} d k^{\prime\prime \, 0} \, \delta_c (k^0 - k^{\prime \, 0}) 
\delta_c (k^0 - k^{\prime\prime \, 0}) G(k^{\prime \, 0}, \Vec{k}) G(k^{\prime\prime \, 0}, \Vec{k}) 
\nonumber\\
&+& {\cal{O}}(f^3) \Big],
\label{C(k)}  
\end{eqnarray}
where we have used the following identity:
\begin{eqnarray}
\delta_c (k^0 - k^{\prime \, 0}) \delta_c (k^{\prime \, 0} - k^{\prime\prime \, 0})
= \delta_c (k^0 - k^{\prime \, 0}) \delta_c (k^0 - k^{\prime\prime \, 0}),
\label{Delta-iden}  
\end{eqnarray}
which can be easily proved.

Moreover, we can rewrite $C(k)$ into a more concise form:
\begin{eqnarray}
C(p) &=& \frac{G(p)}{1 + f \int_C d p^{\prime \, 0} \, \delta_c (p^0 - p^{\prime \, 0}) G(p^{\prime \, 0}, \Vec{p})}
\nonumber\\ \notag 
&=& \frac{G(p)}{1 + f \int_C d^4 k \int_R \frac{d^4 q}{i (2 \pi)^4} \, D_\varphi (k) D_{\varphi^\dagger} (q) 
\delta_c (k^0 + q^0 - p^0) \delta^3 (k + q - p)},
\\[0.5em] 
\label{C(p)}  
\end{eqnarray}
where we have used Eq. (\ref{Def-G}) in the last line. This equation implies that if there were a pole 
at some on-shell value in Eq. (\ref{C(p)}), $p^2 = - m^2$, we would have a bound state with the mass $m$. 
Thus, by solving this equation in the next section, we will investigate the problem of whether there is 
such a bound state or not.

\section{Bound states}

As explained in the previous section, when the pole equation
\begin{eqnarray}
1 + f \int_C d^4 k \int_R \frac{d^4 q}{i (2 \pi)^4} \, D_\varphi (k) D_{\varphi^\dagger} (q) \delta_c (k^0 + q^0 - p^0)
\delta^3 (k + q - p) = 0,
\label{Pole-eq}  
\end{eqnarray}
has a solution at the on-shell value, $p^2 = - m^2$, one can judge that there is a bound state with the mass $m$. Furthermore,
when the bound state has a positive residue at the pole, it is a physical bound state whereas when the bound state has a negative residue, 
it is an unphysical ghost bound state. Anyway, we have to verify that there is a solution to the pole equation (\ref{Pole-eq}).  

For this aim, let us evaluate 
\begin{eqnarray}
J \equiv  \int_C d^4 k \int_R d^4 q \, D_\varphi (k) D_{\varphi^\dagger} (q) \delta_c (k^0 + q^0 - p^0) \delta^3 (k + q - p).
\label{J}  
\end{eqnarray}
Then, the pole equation (\ref{Pole-eq}) is simply written as
\begin{eqnarray}
1 + f \frac{1}{i (2 \pi)^4} J = 0.
\label{J-eq}  
\end{eqnarray}

After integrating over $\Vec{q}$, $J$ is cast to the form:
\begin{eqnarray}
J = \int d^3 k \int_C d k^0 \int_R d q^0 \, \frac{1}{- (k^0)^2 + \omega_k^2} \frac{1}{- (q^0)^2 + (\omega_{p-k}^*)^2}  
\delta_c (k^0 + q^0 - p^0),
\label{J-2}  
\end{eqnarray}
where 
\begin{eqnarray}
\omega_k \equiv \sqrt{\Vec{k}^2 + M^2}, \qquad
\omega_{p-k}^* \equiv \sqrt{(\Vec{p} - \Vec{k})^2 + M^{*2}}. 
\label{2-omega}  
\end{eqnarray}
In evaluating this integral, the point is that $k^0$ is in general complex and takes values along the greatly deformed 
Lee-Wick contour $C$ in the $k^0$-plane, as seen in Fig. 1. This $k^0$-integration along the contour $C$ is then 
reduced to the sum of the integration over the real axis $R$ and residues at two poles, $k^0 = \pm \omega_k$:
\begin{eqnarray}
\int_C d k^0 \rightarrow \left( \int_R + \int_{C(-\omega_k)} - \int_{C(\omega_k)} \right) d k^0,   
\label{C-contour}  
\end{eqnarray}
where $C(\omega)$ denotes an infinitesimal circle rotating anti-clockwise around the pole at $k^0 = \omega$, and the minus sign in front of
the last term on the RHS comes from the clockwise rotation. Consequently, we can obtain
\begin{eqnarray}
&{}& J = \int d^3 k \int_R d q^0 \Bigg[ \int_R d k^0 \frac{1}{- (k^0)^2 + \omega_k^2} \delta (k^0 + q^0 - p^0)
+ \frac{\pi i}{\omega_k} \sum_{\pm} \delta_c (q^0 \pm \omega_k - p^0) \Bigg] 
\nonumber\\
&\times& \frac{1}{- (q^0)^2 + (\omega_{p-k}^*)^2}
\nonumber\\
&=& \int d^3 k \Bigg[  \int_R d k^0 \frac{1}{- (k^0)^2 + \omega_k^2} \frac{1}{- (k^0 - p^0)^2 + (\omega_{p-k}^*)^2}
+ \frac{\pi i}{\omega_k} \sum_{\pm} \int_R d q^0 \delta_c (q^0 \pm \omega_k - p^0) 
\nonumber\\
&\times& \frac{1}{- (q^0)^2 + (\omega_{p-k}^*)^2} \Bigg],
\label{J-3}  
\end{eqnarray}
where we have defined 
\begin{eqnarray}
\sum_{\pm} \delta_c (q^0 \pm \omega_k - p^0) 
= \delta_c (q^0 + \omega_k - p^0) + \delta_c (q^0 - \omega_k - p^0),
\label{sum-pm}  
\end{eqnarray}
and used the fact that $\delta_c (k^0 + q^0 - p^0) \rightarrow \delta (k^0 + q^0 - p^0)$ in the first line of Eq. (\ref{J-3})  
since $k^0, q^0$ and $p^0$ are all real numbers in this case.   

The last term in the second line of Eq. (\ref{J-3}) can be simplified by taking account of a suitable rectangular region whose
form depends on $\delta_c (q^0 - \omega_k - p^0)$ or $\delta_c (q^0 + \omega_k - p^0)$. In case of $\delta_c (q^0 - \omega_k - p^0)$,
the rectangular region is surrounded by the contour which is consisted of the real axis $R$, the finite vertical segment $C_1$ 
located at $q^0 = -\infty$, the horizontal line $R(\omega_k)$ parallel to the real axis 
and running from $-\infty + i {\rm{Im}} \, \omega_k$ to $\infty + i {\rm{Im}} \, \omega_k$
and the finite vertical segment $C_2$ located at $q^0 = + \infty$.  Furthermore, we take this
contour in order not to include the infinitesimal circle rotating anti-clockwise around the pole at $q^0 = - \omega_{p-k}^*$.

\begin{tikzpicture}[>=stealth, scale=1.1]

\draw[->] (-4.5,0) -- (4.5,0) node[below right] {$\mathrm{Re}\,q^0$};
\draw[->] (0,-0.5) -- (0,4) node[above] {$\mathrm{Im}\,q^0$};

\fill (0,0) circle (1pt) node[below right] {$O$};

\coordinate (A) at (-4,0);
\coordinate (B) at (-4,3);
\coordinate (C) at (4,3);
\coordinate (D) at (4,0);

\draw[thick] (A) -- node[left] {$C_1$} (B) -- (C) -- node[right] {$C_2$} (D) -- (A);

\draw[->, thick] (-4,1.2) -- (-4,1.5);

\draw[->, thick] (4,1.5) -- (4,1.2);

\draw[->, thick] (-1,3) -- (-0.5,3);

\draw[->, thick] (-1,0) -- (-0.5,0);

\fill (4,0) circle (1pt);

\node[above] at (-1,3) {$R(\omega_k)$};
\node[above, right] at (0,3.3) {$i\,\mathrm{Im}\,\omega_k$};

\node[below] at (-1,0) {$R$};

\coordinate (wp) at (2,-1);
\fill (wp) circle (2pt) node[right] {$+\omega_{p-k}^*$};

\coordinate (wm) at (-2,1);
\fill (wm) circle (2pt) node[right] {$-\omega_{p-k}^*$};

\draw[thick, ->]
  (-2.5,1) arc[start angle=180, end angle=540, radius=0.5];
  
\draw (-2.1,3) -- (-2.1,1.461);
\draw (-1.9,3) -- (-1.9,1.461);

\draw[->, thick] (-1.8,2) -- (-1.8,2.2);

\draw[->, thick] (-2.2,2.2) -- (-2.2,2);

\draw[thick]
  (A) -- (-2.6,0)
  (-1.4,0) -- (D);

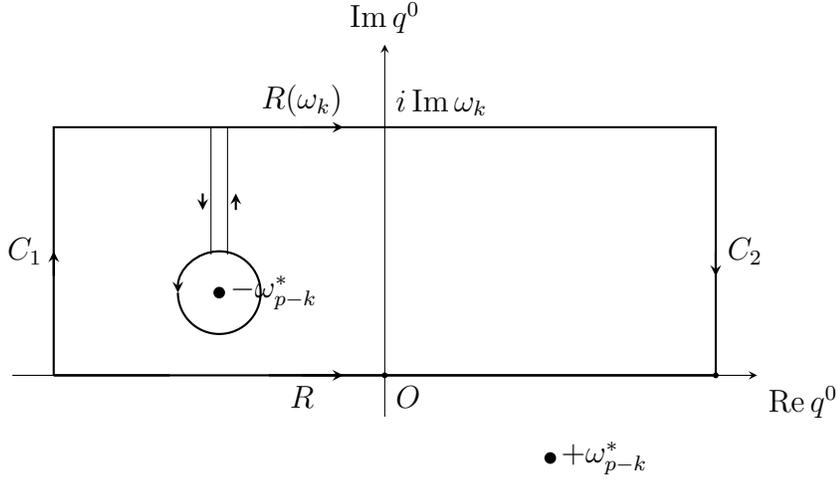
\draw  (0,-1.4) node[below] {Fig.2. The integration contour};

\end{tikzpicture}

Since there is no pole inside such a rectangular region, we have
\begin{eqnarray}
&{}& \int_R d q^0 \delta_c (q^0 - \omega_k - p^0) \frac{1}{- (q^0)^2 + (\omega_{p-k}^*)^2} 
\nonumber\\
&=& \left( \int_{R(\omega_k)} + \int_{C(-\omega_{p-k}^*)} \right)  d q^0 \delta_c (q^0 - \omega_k - p^0) \frac{1}{- (q^0)^2 + (\omega_{p-k}^*)^2},
\label{Rectan}  
\end{eqnarray}
where we have used that the integrals along the finite vertical segments $C_1$ and $C_2$ vanish. The first integral along $R(\omega_k)$
is easily evaluated by making the change of variable from $q^0$ to $q^{\prime \, 0} \equiv q^0 - i \, {\rm{Im}} \, \omega_k$, by which the complex
delta function is reduced to the Dirac delta function, and then the second integral is done by calculating the residue at the pole $q^0 = - \omega_{p-k}^*$. 
\begin{eqnarray}
&{}& \int_R d q^0 \delta_c (q^0 - \omega_k - p^0) \frac{1}{- (q^0)^2 + (\omega_{p-k}^*)^2} 
\nonumber\\
&=& \frac{1}{- (p^0 + \omega_k)^2 + (\omega_{p-k}^*)^2} 
+ \frac{\pi i}{\omega_{p-k}^*} \delta_c [- ( \omega_{p-k}^* + \omega_k ) - p^0].
\label{Rectan1}  
\end{eqnarray}

In a perfectly similar way, one can also calculate the case where $\delta_c (q^0 - \omega_k - p^0)$ is replaced with $\delta_c (q^0 + \omega_k - p^0)$.  
The two results are summarized in
\begin{eqnarray}
&{}& \int_R d q^0 \delta_c (q^0 \pm \omega_k - p^0) \frac{1}{- (q^0)^2 + (\omega_{p-k}^*)^2} 
\nonumber\\
&=& \frac{1}{- (p^0 \mp \omega_k)^2 + (\omega_{p-k}^*)^2} 
+ \frac{\pi i}{\omega_{p-k}^*} \delta_c [\pm ( \omega_{p-k}^* + \omega_k ) - p^0]. 
\label{Rectan2}  
\end{eqnarray}
Accordingly, $J$ reads
\begin{eqnarray}
&{}& J = \int d^3 k \Bigg\{  \int_R d k^0 \frac{1}{k^2 + M^2} \frac{1}{(p - k)^2 + M^{*2}}
+ \frac{\pi i}{\omega_k} \sum_{\pm} \frac{1}{- (p^0 \mp \omega_k)^2 + (\omega_{p-k}^*)^2}
\nonumber\\  
&-& \frac{\pi^2}{\omega_k \omega_{p-k}^*} \sum_{\pm} \delta_c [\pm ( \omega_{p-k}^* + \omega_k ) - p^0] 
\Bigg\},
\label{J-4}  
\end{eqnarray}
Incidentally, it turns out that the first and second terms can be cast to the integral along the Lee-Wick contour $C$:
\begin{eqnarray}
J = \int_C d^4 k \frac{1}{k^2 + M^2} \frac{1}{(p - k)^2 + M^{*2}} \notag
- \int d^3 k \frac{\pi^2}{\omega_k \omega_{p-k}^*} \sum_{\pm} \delta_c [\pm ( \omega_{p-k}^* + \omega_k ) - p^0].
\\[0.5em] 
\label{J-5}  
\end{eqnarray}

Now we are ready to calculate the first term on the RHS of Eq. (\ref{J-4}), which is denoted as $K(p)$ from now on. 
Note that this integral $K(p)$ is a usual one-loop integral except for complex masses, so
we may attempt to make the Wick rotation, by which the $k^0$ integration contour along the real axis is rotated to the
imaginary axis. However, in contrast to the conventional situation, there is a possibility that the contour passes over poles. 
To clarify this situation, we make use of the Feyman's parameter formula and rewrite $K(p)$ as follows:
\begin{eqnarray}
K (p) &\equiv& \int_R d^4 k \, \frac{1}{k^2 + M^2} \frac{1}{ ( p - k )^2 + M^{*2} }
\nonumber\\
&=& \int_R d^4 k  \int_0^1 d x \, \frac{1}{\{ [ k - p (1 - x) ]^2 + p^2 x (1 - x) + M^{*2} + (M^2 - M^{*2}) x \}^2}
\nonumber\\
&=& \int_0^1 d x \int_R d^4 q   \, \frac{1}{ ( q^2 + \Delta )^2},
\label{K-3}  
\end{eqnarray}
where in the last line we have made the change of variables, $q = k - p (1 - x)$ and defined   
\begin{eqnarray}
\Delta \equiv p^2 x (1 - x) + M^{*2} + (M^2 - M^{*2}) x.
\label{Delta}  
\end{eqnarray}
The poles are located at $q^0 = \pm \sqrt{\Vec{q}\,^2 + \Delta}$ in the complex $q^0$-plane. The imaginary part
of the poles comes from the quantity $\Delta$ and is given by
\begin{eqnarray}
{\rm{Im}} \, \Delta = ( 2 x - 1 ) {\rm{Im}} \, M^2,
\label{Im-Delta}  
\end{eqnarray}
which becomes positive for $\frac{1}{2} < x < 1$ due to ${\rm{Im}} \, M^2 > 0$. In this case, since the poles are located
in the first and third quadrants, we have to evaluate the residues from the two poles in making the Wick rotation.\footnote{For 
simplicity, we assume $\Vec{q}\,^2 + {\rm{Re}} \, \Delta > 0$.} After all, the evaluation of contributions from the poles leads to
\begin{eqnarray}
&{}& \int_R d^4 k  \, \frac{1}{k^2 + M^2} \frac{1}{(p - k)^2 + M^{*2}}
\nonumber\\  \notag
&=& \int_I d^4 k  \, \frac{1}{k^2 + M^2} \frac{1}{(p - k)^2 + M^{*2}}
+ \pi i \int_{\frac{1}{2}}^1 d x \int d^3 q \frac{1}{( \Vec{q}\,^2 + \Delta )^{\frac{3}{2}}},
\\[0.5em]
\label{Wick}  
\end{eqnarray}
where $I$ denotes the imaginary axis running from $(0, - i \infty)$ to $(0, i \infty)$ in the complex $k^0$-plane.

Next it is straightforward to calculate the first term on the RHS of Eq. (\ref{Wick}), which is now denoted as $K_I (p)$.\footnote{In Ref. 
\cite{Asorey1}, the authors start with the Euclidean Lagrangian density and calculate only the same quantity as $K_I (p)$ on the basis of 
path integral approach.} As in Eq. (\ref{K-3}), we first rewrite $K_I (p)$ by means of the Feynman parameter formula and then introduce 
the Euclidean momentum which is defined by $q^0 = i k^4$ and $\Vec{q} = \Vec{k}$. Then we have 
$q^2 = - (q^0)^2 + \Vec{q}\,^2 = (k^4)^2 + \Vec{k}^2 \equiv k^2$.  Using these Euclidean momenta, $K_I (p)$ can be recast 
to the form
\begin{eqnarray}
K_I (p) = i \int_0^1 d x \int d^4 k  \, \frac{1}{(k^2 + \Delta)^2}.
\label{K-2}  
\end{eqnarray}

It is obvious that $K_I (p)$ is logarithmically divergent so we adopt the Pauli-Villars regularization. To do that, let us consider 
the following quantity:
\begin{eqnarray}
I (p) = \int d^4 k   \, \left[ \frac{1}{ ( k^2 + \Delta )^2} - \frac{1}{ ( k^2 + \Lambda^2 )^2} \right].
\label{PV}  
\end{eqnarray}
Then, $I (p)$ can be calculated to be
\begin{eqnarray}
I (p) &=& \int d^4 k  \int_\Delta^{\Lambda^2} d y \, \frac{2}{ ( k^2 + y )^3}
\nonumber\\
&=& 4 \pi^2 \int_\Delta^{\Lambda^2} d y \int_0^\infty d k \, \frac{k^3}{ ( k^2 + y )^3} 
\nonumber\\
&=& - \pi^2 \log \frac{\Delta}{\Lambda^2},
\label{PV-2}  
\end{eqnarray}
where we have used the spherical coordinates in the second line. Thus, from  Eqs. (\ref{Delta}) and (\ref{K-2}), 
$K_I (p)$ takes the form:
\begin{eqnarray}
K_I (p) &=& - i \pi^2 \int_0^1 d x \,  \log \frac{p^2 x (1 - x) + M^{*2} + (M^2 - M^{*2}) x}{\Lambda^2}
\nonumber\\
&=& - i \pi^2 \Bigg\{ \log \frac{- M M^*}{\Lambda^2} + \frac{M^2 - M^{*2}}{p^2} \log \frac{M^*}{M}
+ \frac{a(p)}{p^2} 
\nonumber\\
&\times& \log \frac{[ p^2 - M^2 + M^{*2} + a(p) ] [ p^2 + M^2 - M^{*2} + a(p) ]}{4 M M^* p^2} \Bigg\},
\label{K-final}  
\end{eqnarray}
where 
\begin{eqnarray}
a (p) \equiv \sqrt{(p^2 + M^2 - M^{*2})^2 + 4 M^{*2} p^2}.
\label{a(p)}  
\end{eqnarray}

Consequently, the pole equation (\ref{J-eq}), or equivalently (\ref{Pole-eq}), reads
\begin{eqnarray}
&{}& 1 + f \frac{1}{i (2 \pi)^4} \Bigg( - i \pi^2 \Bigg\{ \log \frac{- M M^*}{\Lambda^2} + \frac{M^2 - M^{*2}}{p^2} \log \frac{M^*}{M}
+ \frac{a(p)}{p^2} 
\nonumber\\
&\times& \log \frac{[ p^2 - M^2 + M^{*2} + a(p) ] [ p^2 + M^2 - M^{*2} + a(p) ]}{4 M M^* p^2} \Bigg\}
\nonumber\\
&+& i \pi \int_{\frac{1}{2}}^1 d x \int d^3 q \frac{1}{( \Vec{q}\,^2 + \Delta )^{\frac{3}{2}}}
+ \int d^3 k \frac{\pi i}{\omega_k} \sum_{\pm} \frac{1}{- (p^0 \mp \omega_k)^2 + (\omega_{p-k}^*)^2}
\nonumber\\
&-& \int d^3 k \sum_{\pm} \frac{\pi^2}{\omega_k \omega_{p-k}^*} \delta_c [\pm ( \omega_{p-k}^* + \omega_k ) - p^0] \Bigg) 
\nonumber\\
&=& 0.
\label{Final-pole-eq}  
\end{eqnarray}
Here let us suppose that there is no last term on the LHS, which involves the integration over the complex delta function. 
Then, since the other terms on the LHS have a rather smooth expression with respect to $p^2 (= - m^2)$, 
there could be a possibility such that we have a nontrivial solution to Eq. (\ref{Final-pole-eq}) as suggested 
in Ref. \cite{Asorey1, Asorey2} if the coupling constant $f$ is large enough. However, the presence of the last term 
appears to forbit such a solution owing to the presence of the complex delta function. 

Now we will show that this is not the case. Namely, there is an energy region where the complex delta function
does not contribute, so we can ignore the last term on the LHS in Eq. (\ref{Final-pole-eq}) for this region. 
For our purpose, let us consider a stationary bound state having $\vec{p} = 0$. Then, we have
\begin{eqnarray}
\omega_{p-k}^* + \omega_k = 2 {\rm{Re}} \, \omega_k \ge 2 {\rm{Re}} M, 
\label{Thres}  
\end{eqnarray}
which means that the threshold energy for creating the bound state $\varphi^\dagger \varphi$
is $2 {\rm{Re}} M$ as expected. Note that in this case, the complex delta function reduces to the usual 
Dirac delta function. Using this equation, it turns out that the argument in the delta function does not take 
the vanishing value when $- 2 {\rm{Re}} M < p^0 < 2 {\rm{Re}} M$. Thus, in this energy region, we can
ignore the existence of the complex delta function in Eq. (\ref{Final-pole-eq}).  Moreover, the two terms
just before the last term on the LHS in Eq. (\ref{Final-pole-eq}) involves the integral and is not given by analytic
expressions. These terms might be divergent logarithmically but this diveregence can be absorbed into the cutoff
$\Lambda$ and the remaining part is also smooth with respect to $p^2$. Hence, at least when 
$- 2 {\rm{Re}} M < p^0 < 2 {\rm{Re}} M$ and the coupling constant $f$ is large enough, there is a solution 
satisfying Eq. (\ref{Final-pole-eq}).\footnote{If we use the contour in Ref. \cite{AP} instead of the Lee-Wick's one,
the result might be changed.}

Finally, when there is a bound state in the channel of $\varphi^\dagger \varphi$, the state has the same
norm as the state, $\alpha^\dagger \beta^\dagger | 0 \rangle$. The norm of this state is positive since
we can show that 
\begin{eqnarray}
|| \alpha^\dagger \beta^\dagger | 0 \rangle ||^2 =  \langle 0 | \beta \alpha \alpha^\dagger \beta^\dagger | 0 \rangle
=  \langle 0 | [\beta, \alpha^\dagger] [\alpha, \beta^\dagger] | 0 \rangle = \langle 0 | 0 \rangle = 1,
\label{BS-norm}  
\end{eqnarray}
where we have used the commutation relations (\ref{CRs}) and the definition of the vacuum,
$ \alpha | 0 \rangle = \beta | 0 \rangle = 0$. Thus, the bound state constructed from ghost fields
has the positive norm.

\section{Conclusions}

In this paper, based on the canonical operator formalism we have demonstrated that there is a bound state in the Lee model
which nicely describes some features of QFTs with fourth derivative terms. We should stress that we have only made use of
the operator formalism, which is the basic tool of QFTs. 

Here it is interesting to imagine whether if there were a bound state with positive norm, it could give us a resolution 
to the problem of massive ghost, which violates the unitarity, in the higher derivative 
theories \cite{Stelle, Luca, Anselmi, Salvio, Strumia, Donoghue, Liu, Holdom}. 
We can show that that is not the case since the construction of the bound state does not mean the permanent confinement 
of the massive ghost and the bound state is dissolved into elementary ghost fields in the weak coupling regime. 
To solve the problem of the massive ghost in the higher derivative theories such as quadratic gravity, 
we need the permanent confinement like the confinement of quarks and gluons in QCD.

Under such a situation, it is valuable to pursue a new solution to the massive ghost problem from the viewpoint 
of bound states within the framework of the canonical formalism. For instance, in the analysis of physical modes 
of quadratic gravity \cite{Oda-Can}\footnote{The BRST formalism of various gravitational theories has been already 
constructed in the de Donder (harmonic) gauge \cite{Oda-Q, Oda-W, Oda-Saake, Oda-Corfu, Oda-Ohta, Oda-Conf, Oda-f}.}, 
we usually assume that there is no bound state. If this assumption is not imposed, we are free to take account of a bound state 
which might be constructed from Faddeev-Popov (FP) ghosts and the other elementary fields.  However, if a certain bound state 
is constructed from the massive ghost and BRST FP ghost, and its BRST-conjugate corresponding to this bound state 
is also formed, they together constitute a BRST quartet, thereby belonging to unphysical state and nullifying 
the massive ghost problem \cite{Kawasaki} since this mechanism realizes the scenario of the permanent confinement 
of the massive ghost. Then, an important lesson obtained from the present article is that we should start with 
a higher derivative theory and investigate the bound state problem involving the massive ghost. 
The present formalism is also useful in clarifying the (non-)existence of the bound states. 
We wish to return to this problem in the near future.



\begin{thebibliography}{99}

\bibitem{LW1}
T. D. Lee and G. C. Wick, {``Negative Metric and the Unitarity of the S-matrix'', Nucl. Phys. {\bf B 9} (1969) 209.}

\bibitem{LW2}
T. D. Lee and G. C. Wick, {``Finite Theory of Quantum Electrodynamics'', Phys. Rev. {\bf D 2} (1970) 1033.}

\bibitem{KK1}
J. Kubo and T. Kugo, {"Unitarity Violation in Field Theories of Lee-Wick's Complex Ghost", PTEP {\bf 2023} 
(2023) 123B02.}

\bibitem{KK2}
J. Kubo and T. Kugo, {"Anti-instability of Complex Ghost", PTEP {\bf 2024} (2024) 053B01.}

\bibitem{Asorey1}
M. Asorey, G. Krein, M. Pardina and I. Shapiro, {``Bound States of Massive Complex Ghosts in 
Superrenormalizable Quantum Gravity Theories", JHEP {\bf 01} (2025) 113.}    

\bibitem{Asorey2}
M. Asorey, G. Krein, M. Pardina and I. Shapiro, {``Reflection Positivity in a Higher-derivative Model
with Physical Bound States of Ghosts", arXiv:2511.15283 [hep-th].}    

\bibitem{Lee}
T. D. Lee, {``A Relativistic Complex Pole Model with Indefinite Metric'', in Quanta: Essays in
Theoretical Physics Delicated to Gregor Wentzel, University of Chicago Press, Chicago, 1970, p. 260.}

\bibitem{Nakanishi1}
N. Nakanishi, {``Covariant Formulation of the Complex-Ghost Relativistic Field Theory and the Lorentz 
Noninvariance of the S Matrix'', Phys. Rev. {\bf D 5} (1972) 1968.}

\bibitem{Gell-Low}
M. Gell-Mann and F. Low, {``Bound States in Quantum Field Theory'', Phys. Rev. {\bf 84} (1951) 350.}

\bibitem{Nishijima}
K. Nishijima, {``Formulation of Field Theories of Composite Particles'', Phys. Rev. {\bf 111} (1958) 995.}

\bibitem{Zimmermann1}
W. Zimmermann, {``On the Bound State Problem in  Quantum Field Theory'', Nuovo. Cim. {\bf X} (1958) 597.}

\bibitem{Zimmermann2}
W. Zimmermann, {``Composite Field Operators in  Quantum Field Theory'', Wandering in the Fields: 
Festschrift for Professor Kazuhiko Nishijima on the Occasion of His Sixtieth Birthday, 
World Scientific Publishing, 1987.}

\bibitem{Nakanishi3}
N. Nakanishi, {``A Theory of Closed Unstable Particles'', Prog. Theor. Phys. {\bf 19} (1958) 607.}

\bibitem{Stelle}
K. S. Stelle, {``Renormalization of Higher Derivative Quantum Gravity'',
Phys. Rev. {\bf D 16} (1977) 953.}

\bibitem{Luca}
L. Buoninfante, {``Strict Renormalizability as a Paradigm for Fundamental Physics", 
arXiv:2504.05900 [hep-th] and references therein.}    

\bibitem{Anselmi}
D. Anselmi, {``On the Quantum Field Theory of the Gravitational Interactions", JHEP {\bf 06} 
(2017) 086.}    

\bibitem{Salvio}
A. Salvio, {``Quadratic Gravity'', Front. in Phys. {\bf 6} (2018) 77.}

\bibitem{Strumia}
A. Strumia, {``Interpretation of Quantum Mechanics with Indefinite Norm'', MDPI Physics {\bf 1} (2019) 17.}

\bibitem{Donoghue}
J. E. Donoghue and G. Menezes, {``Unitarity, Stability and Loops of Unstable Ghosts'', Phys. Rev. {\bf D 100} 
(2019) 105006.}

\bibitem{Liu}
J. Liu, L. Modesto and G. Calcagni, {``Quantum Field Theory with Ghost Pairs", JHEP {\bf 02} 
(2023) 140.}    

\bibitem{Holdom}
B. Holdom, {``Making Sense of Ghosts'', Nucl. Phys. {\bf B 1008} (2024) 116696.}

\bibitem{AP}
D. Anselmi, F. Briscese, G. Calcagni, L, Modesto {``Amplitude Prescriptions in Field Theories 
with Complex Poles", JHEP {\bf 05} (2025) 145.}    

\bibitem{Oda-Can}
I. Oda, {``Manifestly Covariant Canonical Formalism of Quadratic Gravity", JCAP {\bf 08} (2025) 079.}    

\bibitem{Oda-Q}
I. Oda, {``Quantum Scale Invariant Gravity in de Donder Gauge'', Phys. Rev. {\bf D 105} (2022) 066001.}   

\bibitem{Oda-W}
I. Oda, {``Quantum Theory of Weyl Invariant Scalar-tensor Gravity'', Phys. Rev. {\bf D 105} (2022) 120618.}   

\bibitem{Oda-Saake}
I. Oda and P. Saake, {"BRST Formalism of Weyl Conformal Gravity", 
Phys. Rev. {\bf D 106} (2022) 106007.}    

\bibitem{Oda-Corfu}
I. Oda, {``BRST formalism of Weyl Invariant Gravity and Confinement of Massive Tensor Ghost'', 
PoS CORFU2023 (2024) 158.}   

\bibitem{Oda-Ohta}
I. Oda and M. Ohta, {``Quantum Conformal Gravity", JHEP {\bf 02} (2024) 213.}    

\bibitem{Oda-Conf}
I. Oda, {``Conformal Symmetry in Quantum Gravity", Eur. Phys. J.  {\bf C 84} (2024) 887.}    

\bibitem{Oda-f}
I. Oda, {``BRST Formalism of $f(R)$ Gravity", Comm. Theor. Phys. {\bf 78} (2026) 025405.}    

\bibitem{Kawasaki}
S. Kawasaki and K. Kimura, {"A Possible Mechanism of Ghost Confinement in a Renormalizable 
Quantum Gravity", Prog. Theor. Phys. {\bf 65} (1981) 1767.}






\end{thebibliography}
\end{document}